\documentclass[fleqn,usenatbib]{mnras}

\usepackage{amsbsy}

\usepackage[T1]{fontenc}
\usepackage{ae,aecompl}
\usepackage{textcomp}
\usepackage{gensymb}

\usepackage{hyperref}
\usepackage{graphicx}	% Including figure files
\usepackage{amsmath}	% Advanced maths commands
\usepackage{changes}    % highlighting of changes LRY
\usepackage{comment}    % commenting large chuncks of text KAP

\usepackage{dirtytalk}

\graphicspath{{pictures/}}

\usepackage{anyfontsize}

\title[Superorbital cycle of HZ~Her/Her~X-1]{
Modeling of 35-day superorbital cycle of $B$ and $V$ light curves of IMXB HZ~Her/Her~X-1}

\author[D. A. Kolesnikov et al]{D. A. Kolesnikov$^{1}$\thanks{E-mail: kolesnikovkda@gmail.com (DAK)},
N. I. Shakura$^{1,2}$,
K. A. Postnov$^{1,2}$,
I. M. Volkov$^{1,3}$,
\newauthor
I. F. Bikmaev$^{2,9}$
T. R. Irsmambetova$^{1}$,
R. Staubert$^{4}$,
J. Wilms$^{5}$,
\newauthor
E. Irtuganov$^{2,9}$,
P. Yu. Golysheva$^{1}$,
S. Yu. Shugarov$^{1,6}$,
I. V. Nikolenko$^{3,7}$,
\newauthor
E. M. Trunkovsky$^{1}$,
G. Sch\"{o}nherr$^{8}$,
A. Schwope$^{8}$,
D. Klochkov$^{4}$
\\
$^1$Sternberg Astronomical Institute, Moscow State University, Moscow, Russia \\
$^2$Kazan Federal University, Kazan, Russia\\
$^3$Institute of Astronomy RAS, Moscow, Russia\\
$^4$Institute for Astronomy and Astrophysics, Tubingen, Germany\\
$^5$Astronomical Institute of the University of Erlangen-Nuremberg, Bamberg, Germany\\
$^6$Astronomical Institute of the Slovak Academy of Scienses, Tatranska Lomnica, Slovakia\\
$^7$Crimean Astrophysical Observatory, Nauchny, Russia\\
$^8$Leibniz Institute for Astrophysics, Potsdam, Germany\\
$^9$Academy of Sciences of Tatarstan, Kazan, Russia\\
}

\date{Accepted XXX. Received YYY; in original form ZZZ}

\pubyear{2019}

\begin{document}

\label{firstpage}
\pagerange{\pageref{firstpage}--\pageref{lastpage}}
\maketitle

\begin{abstract}
The X-ray binary Her X-1 consists of an accreting neutron star and the optical component HZ Her. The 35-day X-ray superorbital variability of this system is known since its discovery in 1972 by the \textit{Uhuru} satellite and is believed to be caused by forced precession of a warped accretion disk tilted to the orbital plane. We argue that the observed features of the 35-day optical variability of HZ Her can be explained by free precession of the neutron star with a period close to that of the forced disk. The model parameters include a) the X-ray luminosity of the neutron star; b) the optical flux from the accretion disk; c) the tilt of the inner and outer edges of the accretion disk. A possible synchronization mechanism based on the coupling between the neutron star free precession and the dynamical action of non-stationary gas streams is discussed.
\end{abstract}


%%%%%%%%%%%%%%%%%%%%%%%%%%%%%%%%%%%%%%%%%%%%%%%%%%

%%%%%%%%%%%%%%%%% BODY OF PAPER %%%%%%%%%%%%%%%%%%

%%\tableofcontents

\section{Introduction}

HZ~Her/Her~X-1 is an intermediate mass X-ray binary consisting of a $1.8-2.0\, M_{\odot}$ evolved sub-giant star and a $1.0-1.5\ M_{\odot}$ neutron star observed as X-ray pulsar \citep{1972ApJ...174L.143T}. The binary orbital period is $P_b=1.7$ days, the X-ray pulsar spin period is $P_x=1.24$ seconds. 
The optical star HZ Her fills its Roche lobe \citep{1972ApJ...178L..65C} to form an accretion disk around the neutron star. 
Before X-ray observations, HZ Her had been classified as an irregular variable. 
Due to the X-ray irradiation, the optical flux from HZ Her is strongly modulated with the orbital period, as was first found by inspecting archive photo plates \citep{1972IBVS..720....1C}. Using the phase connection technique, the timing analysis of  the \textit{RXTE} and \textit{INTEGRAL} observations of Her X-1  enabled the orbital ephemeris of Her X-1 to be updated, the secular decay of the orbital period to be improved, and the orbital eccentricity to be measured  \citep{2009A&A...500..883S}. 

The X-ray light curve of Her X-1 is additionally modulated with an approximately 35-day period \citep{1973ApJ...184..227G}. Most of the 35-day cycles last 20.0, 20.5 or 21.0 orbital periods \citep{1983A&A...117..215S,1998MNRAS.300..992S,Klochkov2006}. The cycle consists of a 7-orbits \say{main-on} state and a 5-orbits \say{short-on} state of lower intensity, separated by 4-orbits intervals during which the X-ray flux vanishes completely. 

Since the first \textit{Uhuru} observations \citep{1973ApJ...184..227G}, the nearly regular 35-day X-ray light curve behaviour of Her X-1 has attracted much attention. 
It is now recognized that the 35-day superorbital cycle of Her X-1 can be explained by the retrograde orbital precession of the accretion disk \citep{1976ApJ...209..562G,1999ApL&C..38..165S}. The 35-d cycle X-ray turn-ons most frequently occur at the orbital phases $\sim 0.2$ or $\sim 0.7$, owing to the tidal nutation of the outer parts of the disk with double orbital frequency when the viewing angle of the outer parts of the disk changes most rapidly \citep{1973NPhS..246...87K,1982ApJ...262..294L,1978pans.proc.....G}. The disk retrograde precession results in consecutive opening and screening of the central X-ray source \citep{1978pans.proc.....G}.
The X-ray light curve is asymmetric between the \say{on} states due to the scattering of X-ray radiation in a hot rarefied corona above the disk. Indeed, the X-ray  \say{turn-on} at the beginning of the \say{main-on} state is accompanied by a significant decrease in the soft X-ray flux because of strong absorption. No essential spectral change during the X-ray flux decrease is observed, suggesting the photon scattering on free electrons of the hot corona close to the disk inner edge \citep{1977ApJ...214..879B,1977MNRAS.178P...1D,1980MNRAS.192..311P,2005A&A...443..753K}.

Soon after the discovery of the X-ray pulsar, the neutron star free precession was suggested as a possible explanation to the observed 35-day modulation \citep{1972Natur.239..325B, 1973AZh....50..459N}. Later on, the EXOSAT observations of the evolution of X-ray pulse profiles of Her X-1 with the 35-day cycle phase was also interpreted by the neutron star free precession \citep{1986ApJ...300L..63T}. \cite{1995pns..book...55S} studied the influence of the free precession of the neutron star to its rotational period. This was further investigated by \cite{2009A&A...494.1025S} and \cite{2013MNRAS.435.1147P} using an extensive set of \textit{RXTE} data. \cite{2013A&A...550A.110S}, however, showed that the possibly existing two \say{35-d clocks}, i.e. the precession of the accretion disk and the precession of the neutron star, are extremely well synchronized -- they show exactly the same irregularity. This requires a strong physical coupling mechanism which could, for example, be provided by the gas-dynamical coupling between the variable X-ray illuminated atmosphere of HZ Her and gas streams forming the outer part of the accretion disk \citep{1999A&A...348..917S,2013A&A...550A.110S}.

The X-ray pulse profile are observed to strongly vary with the 35-day phase \citep{1986ApJ...300L..63T,1998ApJ...502..802D,2000ApJ...539..392S,2013yCat..35500110S} differing significantly at the main-on  and the short-on states. 

Such changes are difficult to explain by the precessing disk only. As was shown by \citealt{2013MNRAS.435.1147P}, the X-ray RXTE/PCA pulse evolution with the 35-day phase could be explained  by the neutron star free precession with a complex magnetic field structure on the neutron star surface. In this model, in addition to the canonical poles (a dipole magnetic field),  arc-like magnetic regions around the magnetic poles are included, which is a consequence of a likely non-dipole surface magnetic field  of the neutron star \citep{1991SvAL...17..339S,1994A&A...286..497P}.

Multiyear X-ray observations shows that there was long (up to 1.5 year) turn-offs of X-ray source, but X-ray irradiation effect was present during these periods  \citep{1985Natur.313..119P, 1994ApJ...436L...9V, 2000ApJ...543..351C, 2004ATel..307....1B, 2004ApJ...606L.135S}. This is probably due to decreasing the angle between the disk and the orbital plane. X-ray source remains obscured by the disk while the angle is close to 0.

Photoplate data shows that there was periods of absence of X-ray irradiation effect \citep{1973ApJ...182L.109J, 1976BAICz..27..325H}. That means accretion vanished during that periods.

Let's return to the 35-day modulation of X-ray flux. In the present paper, we have analyzed extensive optical photometric observations of HZ Her collected from the literature and obtained by the authors. We have found that the model of a precessing tilted accretion disk around a freely precessing neutron star with complex surface magnetic field is able to explain the detailed photometric light curves of HZ Her constructed from all observations available.    

\section{Free precession of neutron star in Her X-1}

%UFN $\rightarrow$ An analysis of X-ray pulse profiles observed by the RXTE (Rossi X-ray Time Explorer) and ‘Ginga’ satellites showed [38] their changing periodically with the same 35-day period as the disk precession.  The neutron star free precession mechanism was suggested as early as in 1972 to explain the 35-day modulation of the observed X-ray flux [39]. 
The free precession occurs when a non-spherical solid body rotates around an axis misaligned with the inertial axes.
Consider two-axial precession of a neutron star rotating with the angular frequency $\omega$, see Fig. \ref{f:NS_free_precession}. If the moments of inertia are $I_1 = I_2, I_3$ and the difference between the moments of inertia is small, $(I_1-I_3)/I_1\ll 1$, the angular velocity of the free precession reads
\begin{equation}
    \Omega_p = \omega\,\frac{I_1 - I_3}{I_1} \cos \gamma
    \label{e:Omegap}
\end{equation}
where $\gamma$ is the angle between the $I_3$ inertia axis and the total angular momentum vector which in this case ($\Omega_p \ll \omega$) almost coincides with the instantaneous spin axis $\omega$.\\ 

From the analysis of the X-ray pulses, in paper \citealt{2013MNRAS.435.1147P} the map of emitting regions on the neutron star surface and the angle between the spin axis and the inertial axis of the neutron star were recovered. The emitting regions include the north and south magnetic poles surrounded by horseshoe-like arcs. The geometry of these regions is due to a complicated non-dipole magnetic field near the neutron star surface. In this model, the emitting arcs enclose the inertial axis, and therefore the storage of accreting matter can produce asymmetry in the principal moments of inertia. In this case, the sign of the precession frequency in equation \ref{e:Omegap} is positive, i.e. the direction of the free precession motion coincides with that of the the neutron star rotation. In the general case, the neutron star can perform a more complicated three-axial free precession \citep{1998A&A...331L..37S}.

The equality of periods of the neutron star free precession and the disk precession is likely to be not coincidental. During the neutron star free precession, the irradiation of the donor star surface strongly changes. The stellar atmosphere heating determines the initial velocity and direction of gas streams flowing through the vicinity of the inner Lagrangian point. In the general case, the gas streams flow off the orbital plane to form the outer parts of a tilted accretion disc. The dynamical action of the streams affects the disk precession, and therefore in such a system, the disk precession can occur synchronously with the neutron star free precession. 

\section{Magnetic forces}

\label{magnetic_forces}
%UFN $\rightarrow$ 
The location of the inner edge of the disk is defined by the break of the disk flow near the magnetospheric boundary at a distance of about 100 neutron star radii ($\sim 10^8$ cm). The magnetic field induces a torque on the inner parts of the disk. %$\leftarrow$ UFN
In the model of interaction of a diamagnetic thin accretion disk with a magnetic dipole \citep{1976PAZh....2..343L,1981AZh....58..765L,1987anz..book.....L}, the magnetic torque averaged over the neutron star spin period reads:
\begin{equation}
\label{e:Km}
  \mathbf{K_m} = \frac{4 \mu^2}{3 \pi R^3_d} \cos \alpha (3 \cos \beta - 1) [\mathbf{n}_{\omega}, \mathbf{n}_{d}].
\end{equation}
Here, $\mu$ is magnetic moment of the neutron star, $R_d$ is inner radius of the disk, $\alpha$ is the angle between the neutron star rotational axis and the inner disk axis, $\beta$ is the angle between the neutron star spin axis and the magnetic dipole. $\mathbf{n}_{\omega}$ is the unity vector along the neutron star spin, $\mathbf{n}_{d}$ is the unity vector along the normal to the inner disk.    

Magnetic torque $K_m$ vanishes if $\alpha = 0\degree$, $\alpha = 90\degree$ or $\beta = \beta_0 = \arccos{\sqrt{3}/3}$. If $\alpha \neq 0^{\degree}$, $\alpha \neq 90\degree$ and $\beta \neq \beta_0$, the magnetic torque is non-zero. The sign of the magnetic torque changes when $\beta$ crosses the critical angle $\beta_0$.

\begin{figure}
    \begin{center}
    \includegraphics[width=\columnwidth]{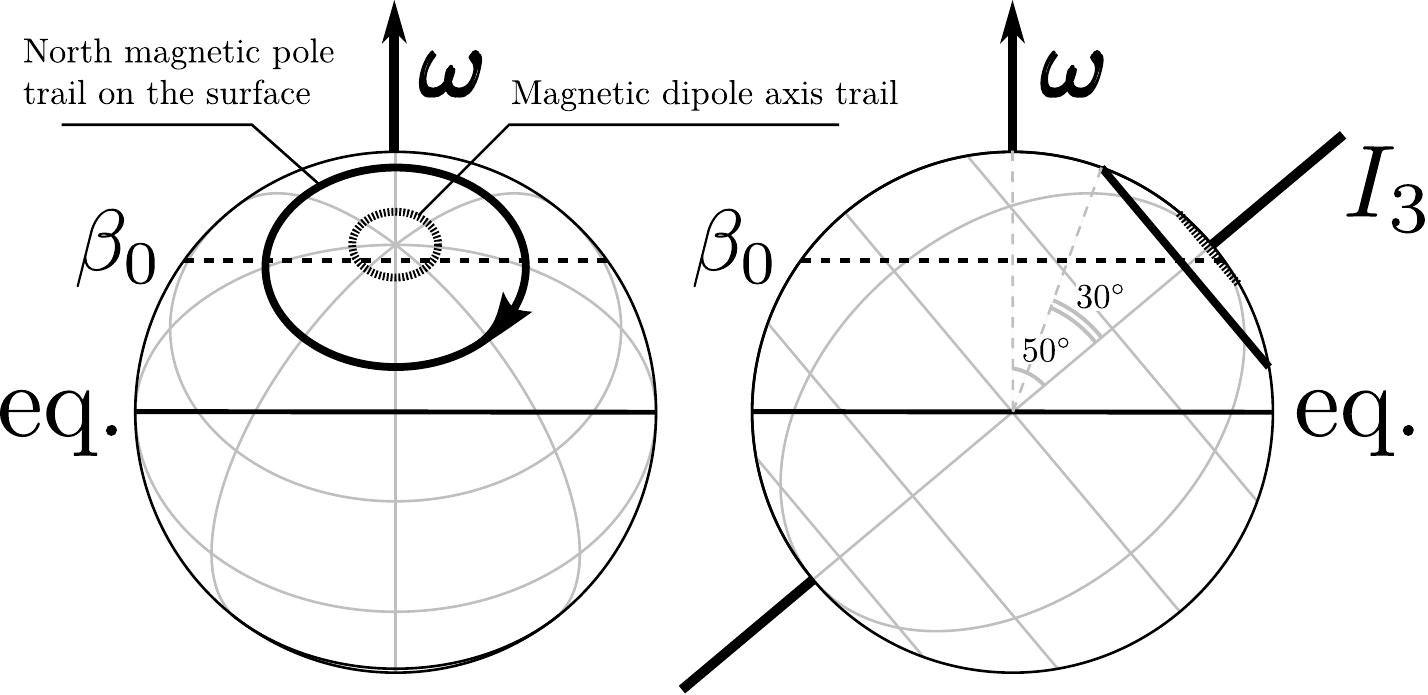}
    \caption{Schematic of the free precession of a neutron star. The angle between the inertia axis $I_3$  and the neutron star spin axis is $50\degree$. The angle between the radius-vector to the north magnetic pole and the $I_3$ axis is $30\degree$. In the course of free precession, the north magnetic pole draws a circle around the $I_3$ axis on the neutron star surface (solid line), and the angle $\beta$ crosses the critical angle $\beta_{0}$ twice. Trail of the magnetic dipole axis on the surface of the neutron star is shown by the smallest circle (gray line). Sign ``eq.'' denotes the equator of the neutron star.}
    \label{f:NS_free_precession}
    \end{center}
\end{figure}

In our model, the angle $\beta$ changes because of the neutron star free precession, see Fig. \ref{f:NS_free_precession}, and the angle $\alpha$ changes because of the disk precession. As a result, the function $K_m(\alpha, \beta)$ must be quite complicated, see Fig. \ref{f:theta_Z}.

A non-zero magnetic torque forces the inner edge of the disk warp with respect to the outer edge. When the magnetic torque vanishes, null warp is expected (the disk becomes flat), which should affect the optical light  curve of HZ Her. Below we construct a geometrical model that takes into account the effect of variable irradiation of HZ Her caused by the warped precessing accretion disk with an account of the 
periodically variable X-ray beam from the freely precessing neutron star. 

\section{\textit{B} and \textit{V} optical observations}\label{s:BV}
To construct optical light curves of HZ Her, the following $B$ and $V$ photometrical observations were used. The 1972 -- 1998 data were compiled from \cite{1973ApJ...181L..39P}, \cite{1972ApJ...177L..97D}, \cite{1973ApJ...186..617B}, \cite{Lyutyj73_PZ}, \cite{1974ApJ...190..365G}, \cite{Lyutyj73_AZH}, \cite{Cherepashhuk_etal74}, \cite{1990PAZh...16..625V}, \cite{1989PAZh...15..806L}, \cite{1978SvAL....4..191K}, \cite{1980PAZh....6..717K}, \cite{1988PAZh...14..438K}, \cite{Kilyachkov94}, \cite{1980A&A....90...54K}, \cite{Gladyshev81}, \cite{1986AZh....63..113M}, \cite{Goransky_Karitskaya86}. In total, the 1972 -- 1998 data include $5771$ individual observations in $B$ and $5333$ in $V$ bands. The 2010 -- 2018 observations were carried out by the authors and include $14034$ $B$ and  $8661$ $V$ individual measurements.

To construct orbital light curves of HZ Her at different phases of 35-day cycle, we have used the orbital ephemeris of the binary system from \cite{2009A&A...500..883S}. The 35-day phases were calculated using X-ray turn-ons of Her X-1 as measured by the \textit{Uhuru}, \textit{Swift}, \textit{RXTE}, \textit{BATSE} (Burst And Transient Experiment) and \textit{INTEGRAL} X-ray observatories, which are kindly provided by R\"{u}diger Staubert. The individual measurement occuring during 35-day cycles with unknown turn-ons have been ignored. The resulted orbital $B$, $V$ light curves in 20 35-day intervals can be found on the GitHub Repository\footnote{ \url{https://github.com/eliseys/data}} and are shown in Fig. \ref{f:phases} by dots.

\section{The model} 
The numerical approach for calculation of the optical light curves of HZ Her is similar to that used by \citet{1971ApJ...166..605W}, \citet{1983MNRAS.202..347H}. Following this method, the surface of the optical star has been split in small areas. Fluxes from the areas visible to the observer sum up to produce the synthetic light curve. It is assumed that the areas have a blackbody spectrum with a temperature defined by the surface local gravity and the X-ray irradiation from the neutron star. Besides, our model implements the complex X-ray shadow formed by the warped disk and non-isotropic X-ray intensity pattern from the neutron star. The C and Python codes of the model are available on the GitHub Repository\footnote{\url{https://github.com/eliseys/discostar}}.

\subsection{Geometry of the donor star}

\begin{figure}
  \includegraphics[width=\columnwidth]{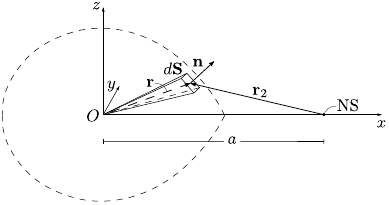}
  \caption{Schematic of the donor star illuminated by the central X-ray source (NS).}
  \label{f:irradiation}
\end{figure}

We assume that the optical star is bounded by an equipotential surface corresponding to the Roche potential. In the Cartesian coordinates $xyz$ with the origin at the center of mass of the optical star which is rotating synchronously with the binary system, the Roche potential is \citep{1959cbs..book.....K}:
\begin{equation}
  \label{e:roshe_potential}
  \psi = \frac{G m_1}{r_1} + \frac{G m_2}{r_2} + \frac{\omega^2}{2} \left[ \left( x - \frac{m_2 a}{m_1+m_2} \right)^2 + y^2 \right] \, ,
\end{equation}
where $xy$ axes are in the orbital plane, $x$ axis is forwarded to secondary star center, $r_1 = \sqrt{x^2+y^2+z^2}$ and $r_2 = \sqrt{(a - x)^2+y^2+z^2}$ is the distance from the center of mass of the first star and the second star to a given point, respectively, $a$ is the distance between centers of mass of the stars, see Fig.~\ref{f:irradiation}.

In the dimensionless units ($a = 1$), the Roche potential reads (see, e.g., \cite{Cherep2013}):
\begin{multline}
  \Omega = \frac{\psi}{G m_1} - \frac{m_2^2}{2 m_1 (m_1 + m_2)} = \frac{1}{r} + q \left( \frac{1}{\sqrt{1-2x+r^2}} - x \right) + \\ 
  + \frac{1}{2}(1+q)(x^2+y^2) \, ,
\end{multline}
where $r = \sqrt{x^2 + y^2 + z^2}$ is the distance from the center of mass of the optical star $m_1$, $q = m_2/m_1$ is the compact to optical component mass ratio.

The value of $\Omega$ is defined through the Roche lobe fill fraction:  
\begin{equation}
  \label{mu}
  \mu = \frac{R_0^\star}{R_0} \, ,
\end{equation}
where $R_0$ is the polar radius of the Roche lobe,  $R_0^{\star}$ is the polar radius of the star. Two parameters $q$ and $\mu$ explicitly define the shape of the donor star. 

The unit normal vector to a point on the surface of the star is defined by the Roche potential gradient:
\begin{equation}
  \label{n}
  \mathbf{n} = \frac{\nabla \Omega}{|\nabla \Omega|} \, .
\end{equation}

The gravity acceleration vector is:
\begin{equation}
  \label{g}
  \mathbf{g} = - \nabla \Omega \, .
\end{equation}
The vector of a surface element in the spherical coordinate system with the origin in the star's barycenter is:
\begin{equation}
  \label{dS}
  d\mathbf{S} = \frac{\mathbf{n}\,dS}{\mathbf{n}\cdot\mathbf{r}}
\end{equation}
where $dS = r^2\,d\varphi\,d\theta\cos\theta$ is the  surface area element; the factor $1/(\mathbf{n}\cdot\mathbf{r})$ is introduced because the surface of the star is not perpendicular to the vector $\mathbf{r}$, see Fig.~\ref{f:irradiation}.

The projection of the surface element vector onto the sky plane is: 
\begin{equation}
  \label{dS_projection}
  \mathbf{n}_o \cdot d\mathbf{S} \, ,
\end{equation}
where $\mathbf{n}_o$ is the unity vector pointed to the observer. 

\subsection{Surface temperature of the donor star}
\label{surface_temperature}

We assume that the donor star surface at any point emits a blackbody spectrum. The radiation flux from the blackbody surface element $dS$ is defined by its  temperature $T$:
\begin{equation}
  dF = \sigma_B T^4 dS\,,
\end{equation}
where $\sigma_B$ is the Stefan-Boltzmann constant. Due to the temperature variation over the stellar surface, different parts of the star differently contribute to the total flux. The gravitational darkening and X-ray irradiation are also taking into account in the surface temperature change. 

The gravitational darkening depends on the surface gravity $g$: 
\begin{equation}
  T = T_0 \left( \frac{g}{g_0} \right)^{\beta} \, , 
\end{equation}
where $T_0$ and $g_0$ are the temperature and gravity acceleration at the donor star pole, respectively. The coefficient $\beta$ is set to $0.08$ \citep{1924MNRAS..84..665V}. The polar temperature $T_0$ is free parameter, and $g$ is calculated by differentiating equation \ref{e:roshe_potential}.

In the presence of X-ray irradiation, the surface element $dS$ is illuminated by the external radiation flux $A dF_x$, where $A$ is the fraction of the thermalised X-ray flux. Thus, the total radiation flux from the surface element reads:
\begin{equation}
    \sigma T_{irr}^4 dS = dF + A dF_x \, ,
\end{equation}
where $T_{irr}$ is the effective temperature of the element with an account of the X-ray irradiation.

\subsection{Geometry of the accretion disk}

Here we introduce a formalism which enables us to calculate the X-ray shadow produced by a warped, inclined disk with a finite width of the outer edge. 

The accretion disk is modelled by $N$ circle rings and an external cylindrical belt centered on the neutron star. The orientation of the each individual ring $i$ is determined by the orthonormal vector $\mathbf{d}_{i}$, $i = \{1,2\dots\,N\}$. The external cylindrical belt is described by the radius $R$ and height $H$.

We assume that the disk is mostly  warped  near its inner edge because of the interaction with the neutron star magnetosphere. The radii of the circles representing such a disk are ordered as $r_1 \gg r_{2} > r_{3} > \, \ldots \, > r_N$, where $r_1$ is the radius of the outermost ring and the external belt, $r_N$ is the radius of the innermost ring. 
\begin{figure}
  \begin{center}
    \includegraphics[width=\columnwidth]{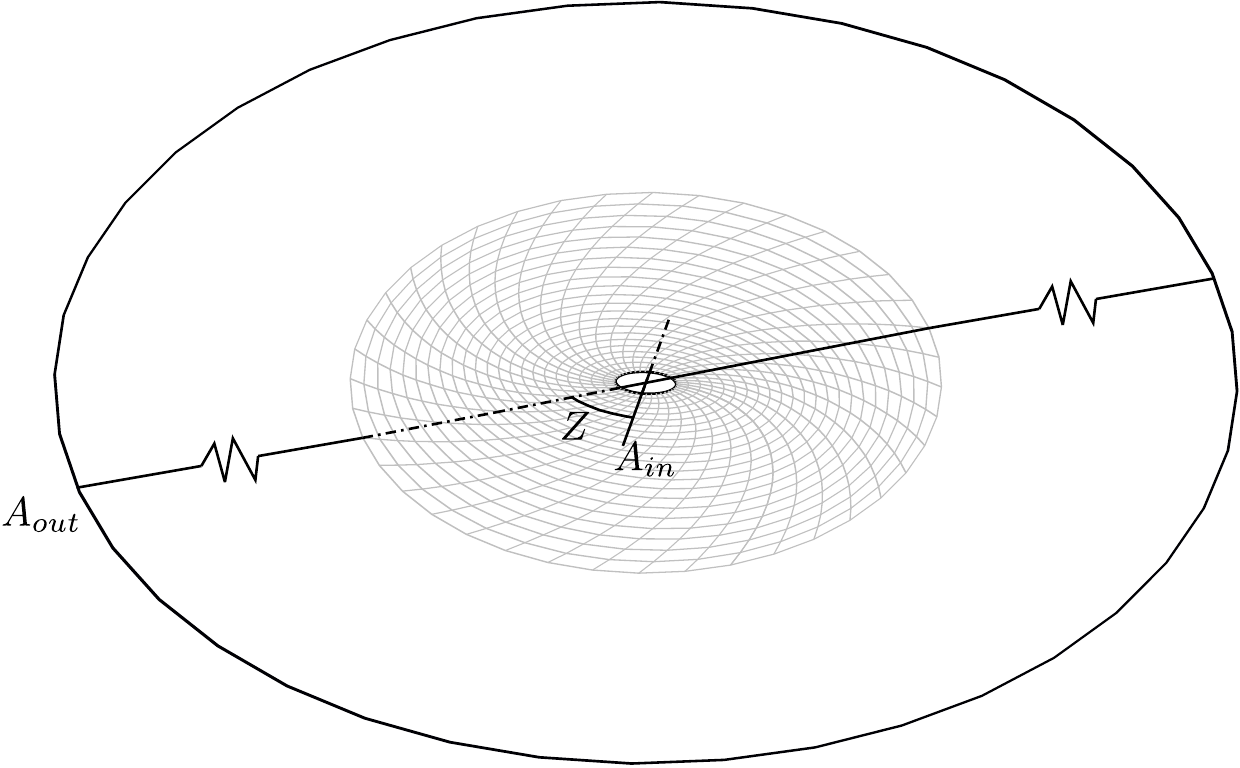}
    \caption{Model of the disk. Disk is modelled by $N$ circle rings. Central part is shown not to scale. Radius of the outer ring is much larger than radii of rest of the rings: $r_1 \gg r_2 > r_3 \dots r_N$. Outer and inner circle has the nodal lines $A_{out}$ and $A_{in}$ respectively. The angle between $A_{out}$ and $A_{in}$ is the twist angle $Z$.}
    \label{f:disk_Z}
  \end{center}
\end{figure}
In our model, we assume that the coordinates $\{\theta_i, \varphi_i\}$ of the vector $\mathbf{d}_{i}$ change linearly with the index $i$:
\begin{equation}
  \theta_i = \theta_{1} + i \, \frac{\theta_{N} - \theta_{1}}{N-1}\,,
\end{equation}
\begin{equation}
  \varphi_i = \varphi_{1} + i \, \frac{\varphi_{N} - \varphi_{1}}{N-1}\,.
\end{equation}
This enables us not to define the position of each ring but only of the innermost $\mathbf{d}_{N} \equiv \mathbf{d}_{in}$ and the outermost $\mathbf{d}_{1} \equiv \mathbf{d}_{out}$ rings.

To describe the disk twist, we introduce the twist angle $Z \equiv \varphi_{out} - \varphi_{in}$ between the nodal lines of the outermost and innermost rings, see Fig. \ref{f:disk_Z}.  

The X-ray radiation flux from the neutron star passing between the $i$-th and the $i+1$-th ring is blocked. The cylindrical belt also screens the X-ray radiation from the central neutron star within the solid angle $2\pi H/R$.

%% It means that the disk is warped mostly near the neutron star magnetosphere where the magnetic torque is applied. This implies that on scales comparable to $R$ the disk can be considered flat, which simplifies the calculation of the disk shadow on the optical star surface in the course of the disk precessional motion. 

\subsection{Transit and eclipse of the accretion disk}

During the orbital motion, the accretion disk and the X-ray source are eclipsed by the donor star near the orbital phase $0$. Oppositely, the disk and the X-ray source transits in front of the donor star near the orbital phase $0.5$. This gives rise to the main and secondary minima on the light curves at the orbital phases $0$ and $0.5$, respectively.   

In the present study, we have not modelled the primary minima of the orbital light curves and have excluded the orbital phases 0.0--0.13 and 0.87--1.0 from calculations of the synthetic orbital light curves.
To model the secondary minimum of the orbital light curves we have used the ray-marching technique.

%%At the orbital phases around 0.0, the optical star HZ Her passes between the observer and the X-ray source surrounded by the accretion disk. In order to calculate the optical light curves from HZ Her at these phases, we need to model the brightness distribution over the disk surface. For a twisted accretion disk it is a difficult task which we postpone for future studies.

The contibution from the disk $F_d$ to the observed light curve is defined by the dimensionless parameters $F_B$ and $F_V$: 
%To account for the contribution from the disk to the observed total flux, we introduced parameters $F_B$ and $F_V$ 
%corresponding to the Johnson-Cousins $B$ and $V$ normalized specific fluxes from the disk: 
\begin{equation}
    F_{B,V} = \left( \frac{F_d}{F_0} \right)_{B,V} \,.
\end{equation}
Here $F_0$ is the flux from the optical star at the orbital phase $0$. The parameters $F_B$ and $F_V$ are assumed to be constant during the orbital period but can vary with the  35-day phase. 

\subsection{X-ray emission from the neutron star}

\begin{figure}
  \begin{center}
    \includegraphics[width=\columnwidth]{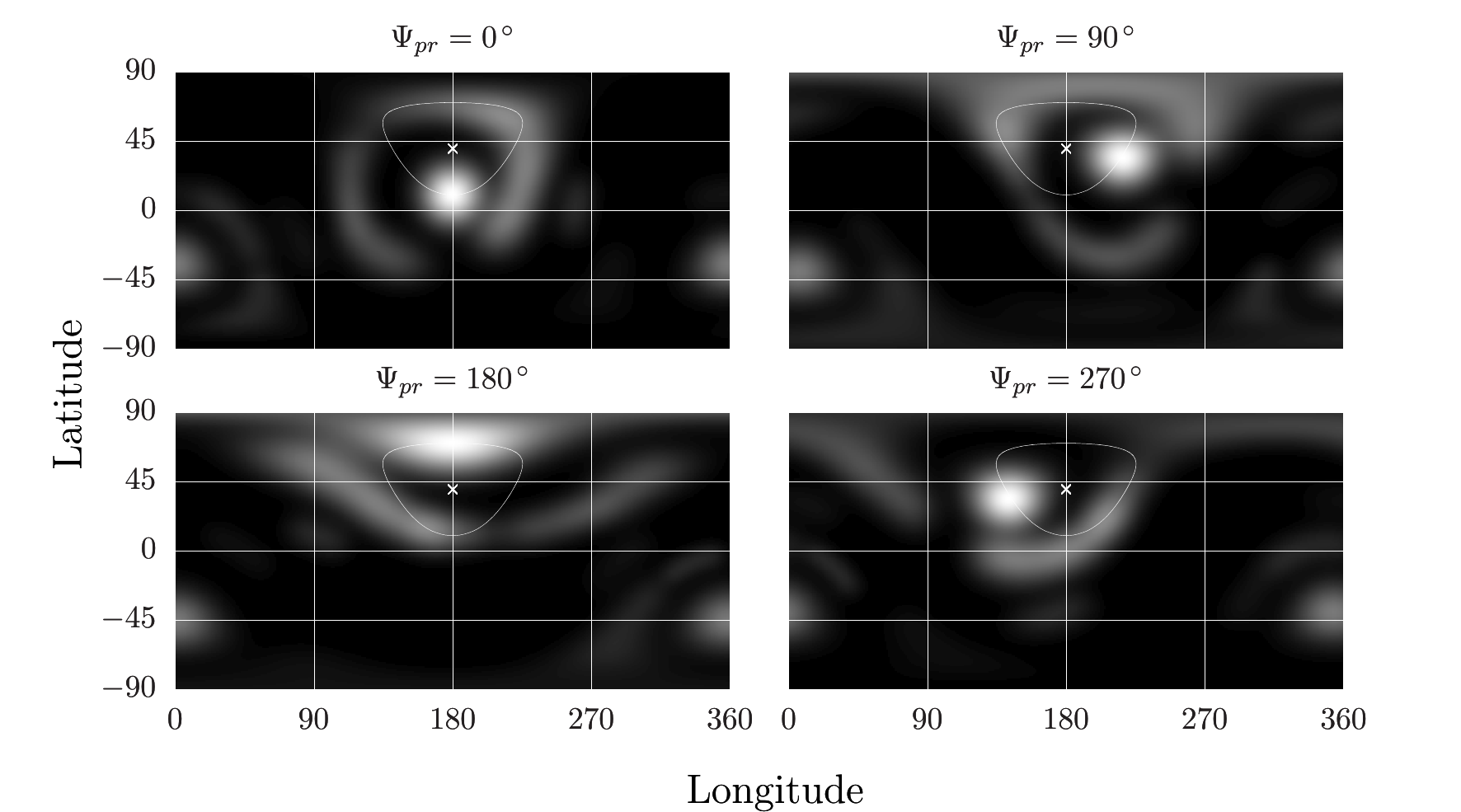}
    \caption{X-ray intensity from the neutron star surface in Her X-1 as function of spherical coordinates at the free precession phases $\Psi=0.0$, $0.25$, $0.5$ and $0.75$ \citep{2013MNRAS.435.1147P}. The latitude of the neutron star's north pole is $90\degree$. The longitude $180\degree$ corresponds to the meridian passing through the poles and the magnetic dipole at the free precession phase $\Psi=0$. This figure was produced by summation of intensity of all emitting regions on the surface of the neutron star. White ring here is the same as solid ring on the Fig.~\ref{f:NS_free_precession}}
    \label{f:NS_map}
  \end{center}
\end{figure}

To calculate the temperature distribution over the surface of the donor star illuminated by X-rays from the neutron star, we have used non-isotropic X-ray source intensity which was adopted from the model constructed by \cite{2013MNRAS.435.1147P}. The X-ray intensity pattern was derived from the analysis of X-ray pulse evolution of Her~X-1 with the 35-day period. In this model, the neutron star experiences the free precession with a period close to 35 days, see Fig.~\ref{f:NS_map}. According to the model, X-ray emission leaves the neutron star perpendicular to its surface in \say{narrow pencil beams}, such that one does not have to worry about gravitational bending. The X-ray diagram illuminating the optical star HZ~Her is obtained by averaging the emission from the neutron star surface over the fast neutron star spin period $1.24\,\mathrm{s}$. The averaged intensity is modulated by the neutron star free precession.

\begin{figure}
    \begin{center}
    \includegraphics[width=0.5\columnwidth]{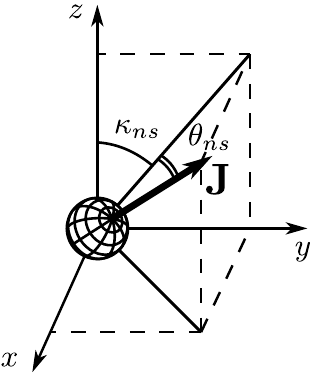}
    \caption{
        Orientation of the neutron star angular momentum $\mathbf{J}$ relative to the observer. The origin of the coordinate system is at the neutron star center. The axes $x$ and $z$ lie in the sky plane. The axis $x$ is pointed to the observer. The axis $z$ goes along the projection of the binary orbital momentum vector to the sky plane. The angle $\theta_{ns}$ is the angle between $\mathbf{J}$ and the sky plane. The angle $\kappa_{ns}$ is the angle between the projection of $\mathbf{J}$ on the sky plane and the $z$ axis.
    }
    \label{f:kappa_theta}
    \end{center}
\end{figure}

Following \cite{2013MNRAS.435.1147P} we set the angle $\theta_{ns}$ between the rotation axis of the neutron star and the sky plane to $-3\degree$. The minus sign means that the north hemisphere is directed off the observer. The second angle $\kappa_{ns}$, which is the position angle of the rotation axis of the neutron star, is a free parameter, see Fig. \ref{f:kappa_theta}. 

In the case of non-isotropic X-ray emission from the neutron star the irradiation flux impinging a surface elemental area of the optical star is  
\begin{equation}
  \label{dF_non_iso}
  dF_{\mathrm{x}} = I_{\mathrm{x}}(\mathbf{r}_2) \, \frac{d\mathbf{S} \cdot \mathbf{r}_2}{r_2} \, ,
\end{equation}
where $I_{\mathrm{x}}(\mathbf{r}_2)=dL_x(\mathbf{r}_2)/d\Omega$ is the X-ray intensity in the direction $\mathbf{r}_2$ (see Fig. \ref{f:irradiation}).

%% \begin{table}
%%   \caption{Free parameters of the model}
%%   \label{tab:free_parameters}
%%   \begin{tabular}{l|c}
%%     \hline
%%     \hline
%%     Parameter & Designation \\
%%     \hline
%%     Semi-major axis                           &   $a$ \\
%%     Mass ratio, $M_x/M_v$                     &   $q$ \\
%%     Roche lobe filling factor                 &   $\mu$ \\
%%     Gravity darkening coefficient             &   $\beta$ \\
%%     X-ray reprocessing factor                 &   $A$ \\
%%     Disk's radius                             &   $R$ \\
%%     Disk's width                              &   $h$ \\
%%     NS's orientation angle                    &   $\kappa_{ns}$ \\
%%     NS's orientation angle                    &   $\theta_{ns}$ \\
%%     NS's zero-phase of free precession        &   $\Psi_{0}$ \\
%%     Star's polar temperature                  &   $T_0$ \\
%%     Inclination                               &   $i$ \\
%%     Disk's max opening phase                  &   $\Phi_0$ \\
%%     NS X-ray luminosity                       &   $L_X$ \\
%%     Disk outer edge tilt                      &   $\theta_{out}$ \\
%%     Disk inner edge tilt                      &   $\theta_{in}$ \\
%%     Disk normalized $B$-flux                  &   $F_B$ \\
%%     Disk normalized $V$-flux                  &   $F_V$ \\
%%     Disk twist, $\varphi_{out}-\varphi_{in}$  &   $Z$ \\
%%     \hline
%%     \hline
%%   \end{tabular}
%% \end{table}

\section{Modelling}
\label{modelling}
We assume that the retrograde disk precession phase $\Phi$ changes linearly with time. The phase angle of the outer disk edge is defined as a function of the 35-day phase $n = \{0,1,2\dots 19\}$ 
\begin{equation}
\label{e:Phi}
  \Phi = - \, n \frac{2\pi}{N} + \Phi_0\,,
\end{equation}
where $N = 20$ is the number of discrete phases of the 35-day cycle. The phase angle $\Phi_0$ is the initial disk precession angle. At the phase angle $\Phi_0$  the disk is mostly opened to the observer. The value of $\Phi_0$ is treated as free parameter.

The neutron star free precession is assumed to be a prograde linear function of $n$ with the initial phase angle $\Psi_0$:
\begin{equation}
  \Psi = n \frac{2\pi}{N} - \Psi_0\,.
\end{equation}
$\Psi_0$ is the phase angle at which the north magnetic pole of the neutron star passes most closely to the neutron star equator. Following \cite{2013MNRAS.435.1147P} we set $\Psi_0=2\pi/20$.

\subsection{Free parameters of the model}

The model parameters are summarized in Tables \ref{tab:parameters1} and \ref{tab:parameters2}. They list the parameters that have been fixed and changing during the 35-day cycles, respectively. The fixed model parameters have been taken from the literature, and the changing parameters have been found by best-fitting the model optical light curves within the limits shown in the second column of Table \ref{tab:parameters2}. 

\begin{table*}
  \caption{The model parameters fixed during the $35^d$ cycle}
  \label{tab:parameters1}
  \begin{tabular}{l|l|l|l}
    \hline
    \hline
    Parameter & Symbol & Value & Ref.\\
    \hline
    Semi-major axis     & $a$                     &   $6.502 \times 10^{11}$ cm & \cite{2014ApJ...793...79L} \\
    Mass ratio, $M_x/M_v$ & $q$               &   $0.6448$ & \cite{2014ApJ...793...79L}\\
    Roche lobe filling factor & $\mu$              &   $1.0$ & assumed\\
    Gravity darkening coefficient & $\beta$    &   $0.08$ & assumed\\
    X-ray reprocessing factor &  $A$            &   $0.5$ & assumed \\
    Disk radius & $R/a$                    &   $0.24$ & assumed\\
    Outer disk thickness & $H/R$  &   $0.15$ & assumed \\
    NS orientation angle & $\kappa_{ns}$   &   $10\degree$ & assumed\\
    NS orientation angle & $\theta_{ns}$   &   $-3\degree$ & \cite{2013MNRAS.435.1147P} \\
    NS initial phase angle  &   $\Psi_0$        &   $2\pi/20$ & \cite{2013MNRAS.435.1147P}\\
    Star's polar temperature & $T_0$         &   $7794.0$ K & \cite{2014ApJ...793...79L}\\
    Binary inclination & $i$                        &   $88.93\degree$ & calculated to reproduce correct main-on/short-on beginning\\
    Disk max opening phase angle &      $\Phi_0$        &   $2\pi/5$& assumed \\
    \hline
    \hline
  \end{tabular}
\end{table*}

\begin{table}
  \caption{The model parameters changing with the $35^d$ cycle phase}
  \label{tab:parameters2}
  \begin{tabular}{l|l|l}
    \hline
    \hline
    Parameter & Symbol & Limits \\
    \hline
    NS X-ray luminosity & $L_x$                &   $0.1\,...\,10 \times 10^{37}$ erg/s\\
    Disk outer edge tilt & $\theta_{out}$      &   $0\,...\,40\degree$\\
    Disk inner edge tilt & $\theta_{in}$       &   $0\,...\,40\degree$\\
    Disk normalized $B$-flux  &  $F_B$               &   $0\,...\,4$\\
    Disk normalized $V$-flux  &  $F_V$               &   $0\,...\,4$\\
    Disk phase angle & $\Phi$ & $-20\degree \dots 20\degree$ \\
    && dev. from linear law\\
    Disk twist $\varphi_{out}-\varphi_{in}$ & $Z$ &    $-90\,...\,+90\degree$\\
    && with $10\degree$ step\\
    \hline
    \hline
  \end{tabular}
\end{table}

\subsection{Results of the modelling}

The results of best-fitting of the model parameters  with observed $B$ and $V$ light curves constructed for 20 intervals of the 35-day cycle are presented in Figures \ref{f:theta_Z}, \ref{f:F}, \ref{f:Lx}, \ref{f:epsilon}, \ref{f:phi}. The observed $B$ and $V$ light curves constructed for 20 intervals of the 35-day cycle according to the procedure described above in Section \ref{s:BV} are presented in Fig. \ref{f:phases}. 

\subsubsection{Disk tilt angles}

\begin{figure}
    \centering
    \includegraphics[width=0.45\textwidth]{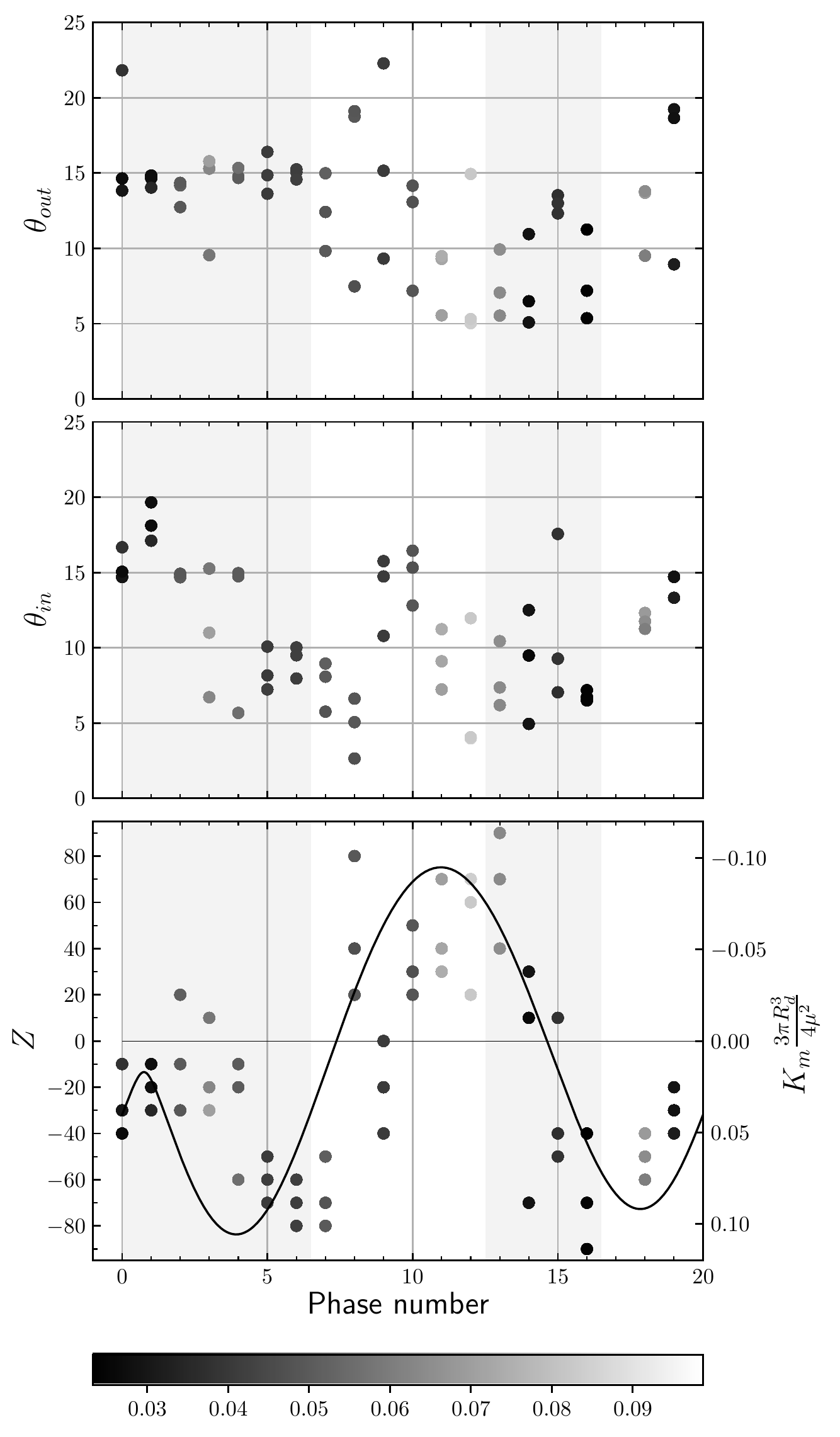}
    \caption{ 
    %Disk's warp angle $Z$, tilt of the inner edge $\theta_{in}$, tilt of the outer edge $\theta_{out}$.
    Disk tilt angles to the orbital plane $\theta_{in}, \theta_{out}$ and the disk twist angle $Z$ as a function of the 35-day cycle phase number $n$ (the upper, middle and bottom panels, respectively). The solid line in the bottom panel shows the expected magnetic torque $K_m$ acting on the inner disk (in dimensionless units, right axis) from the freely precessing neutron star with param-eters from Table \ref{tab:parameters1}. The gray vertical strips mark the main-on and short-on states of Her X-1. The filled circles gray-coloured with reduced $\chi^2$ values (the scale in the bottom) correspond to three best-fit model light curves presented in Fig. \ref{f:phases}. 
    }
    \label{f:theta_Z}
\end{figure}

\begin{figure*}
  \begin{center}
    \includegraphics[width=\textwidth]{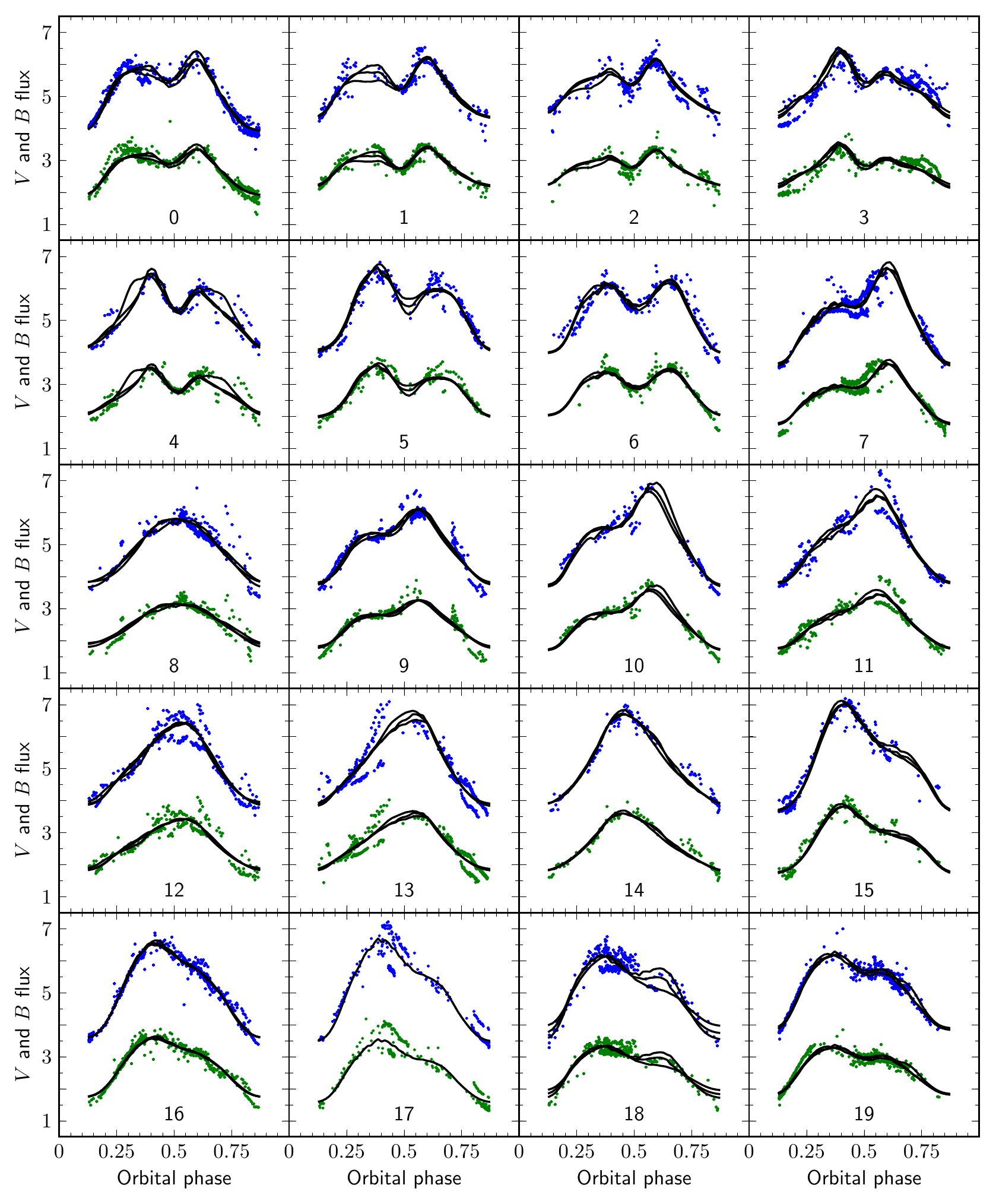}
    \caption{Orbital $B$, $V$ light curves constructed in 20 phase intervals of the 35-day cycle. $B$ points are shifted by 2 units for clarity. The fluxes are normalized to the primary minimum. The 35-day intervals are marked with numbers $n$ according to equation \ref{e:Phi}. The solid curves show the best-fit models  calculated for the grid of the twist angle $Z$ as shown in the bottom panel in Fig. \ref{f:theta_Z}.}
    \label{f:phases}
  \end{center}
\end{figure*}

In Fig. \ref{f:theta_Z} we show the disk tilt angles to the orbital plane (the inner and outer disk tilts $\theta_{in}, \theta_{out}$) and the disk twist angle $Z$ as a function of the 35-day cycle phase number $n$, for three best-fit models.  The model light curves were calculated for an even grid of $Z$ from $-90\degree$ to $+90\degree$ with $10\degree$ step.
Three model light curves with $Z$ producing the minimal reduced $\chi^2$ values are shown in 
Fig. \ref{f:phases} by solid lines. 
The $\chi^2$ is calculated as 
\begin{equation}
    \chi^2 = \frac{1}{N - N_{var}} \sum_i^N (y_i - f(x_i))^2 \,
\end{equation}
where $N$ is the number of observed points in each 35-day phase interval (typically several hundreds), $N_{var}=6$ is the number of fitting parameters for each twist angle $Z$ (see Table \ref{tab:parameters2}), $y_i$ and $x_i$ is the observed flux and orbital phase, respectively, $f(x)$ is the synthetic light curve.

The filled circles Fig.~\ref{f:theta_Z} are $\chi^2$ gray-coloured  (the scale at the bottom of the Figure) and correspond to three best-fit light curves from Fig. \ref{f:phases}. The phase interval marked with $n=0$ corresponds to the beginning of the main-on state. The light gray vertical strips mark the main-on and short-on states of Her X-1. 

Fig. \ref{f:theta_Z} suggests that the outer disk tilt $\theta_{out}$ (the upper panel of the Figure) stays at about $15\degree$  during the main-on and tends to lower values about $10\degree$ during the short-on. The inner disk tilt $\theta_{in}$ (the middle panel)  varies between \textbf{$\sim 15\degree$} and $\sim 5\degree$. The $Z$-angle describing the disk twist (the lower panel of the Figure) strongly changes between $\sim -90\degree$ and $\sim +90\degree$. In our model, zero twist angle corresponds to null magnetic force moment $K_m$ (the right scale of the bottom panel) when the accretion disk is the least warped. The solid line in the bottom panel shows the expected magnetic torque acting on the inner edge of the disk see equation \ref{e:Km} with the adopted fixed parameters of the neutron star shown in the Table \ref{tab:parameters1}.
Therefore, the change of the parameters of the warped twisted accretion disk shown in Fig. \ref{f:theta_Z} with the 35-day phase are in qualitative agreement with our physical model.  

\subsubsection{Disk fluxes}

\begin{figure}
    \centering
    \includegraphics[width=0.4\textwidth]{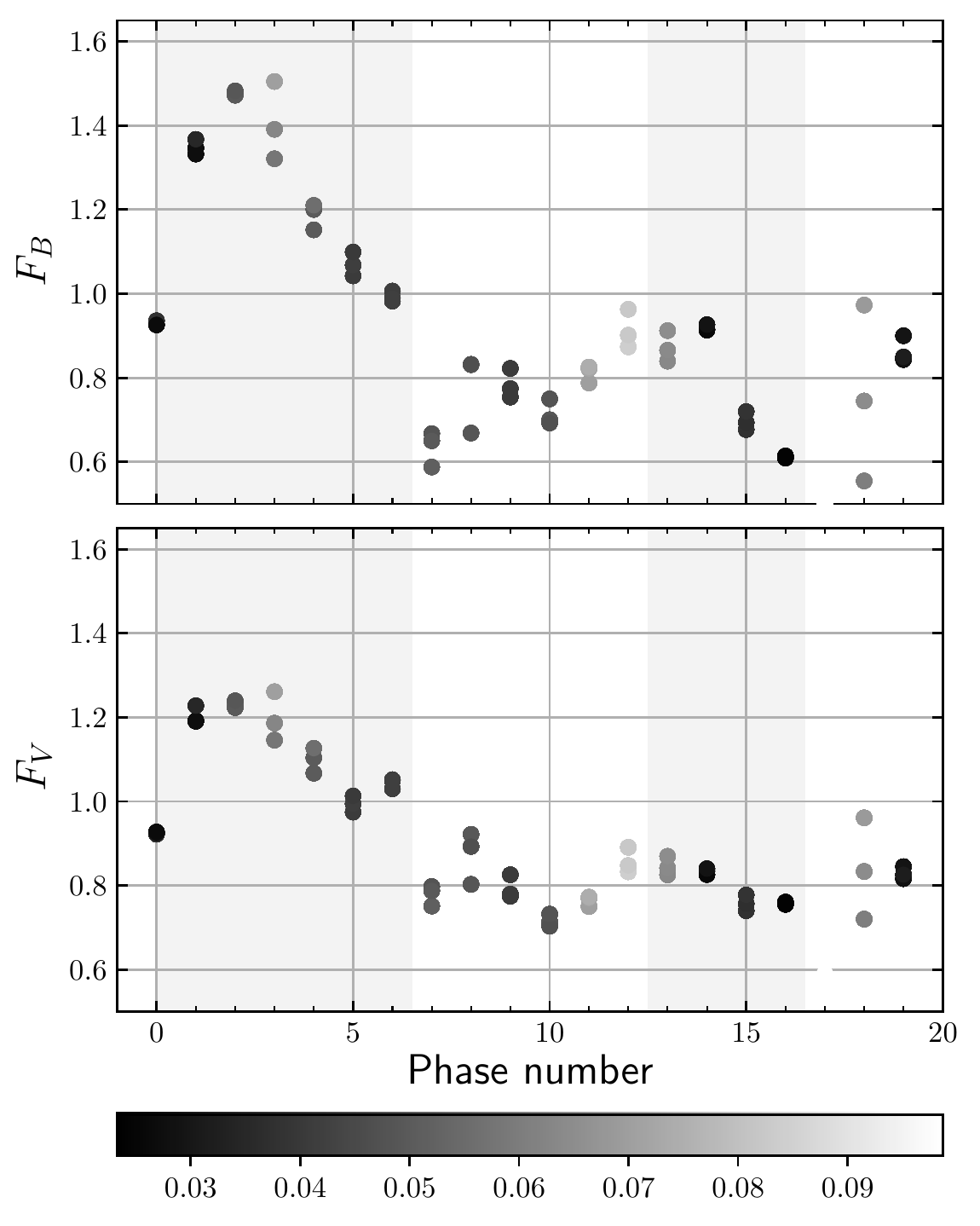}
    \caption{The proper optical B,V-fluxes from the disk for the models described in Fig. \ref{f:theta_Z} normalized to the optical star flux at the primary eclipse.}
    \label{f:F}
\end{figure}

Fig. \ref{f:F} shows the proper disk fluxes $F_B$ (the upper panel) and $F_V$ (the lower panel) in units of the optical $B$, $V$ fluxes from the star at the orbital phase 0 (the primary orbital minimum). The filled gray-scaled circles show the same models as in Fig. \ref{f:theta_Z}. It is seen that the disk flux is maximum during the main-on state and points to the existence of the second maximum at 35-day phases 0.65--0.70 at the beginning of the short-on. The first maximum is higher because of the additional (on top of purely geometrical view) irradiation of the outer parts of the disk by the central X-ray source. Such a behaviour of the disk flux is expected in our model because the zero phase of the neutron star free precession is close to the beginning of the main-on. 

\subsubsection{X-ray luminosity}

\begin{figure}
    \centering
    \includegraphics[width=0.4\textwidth]{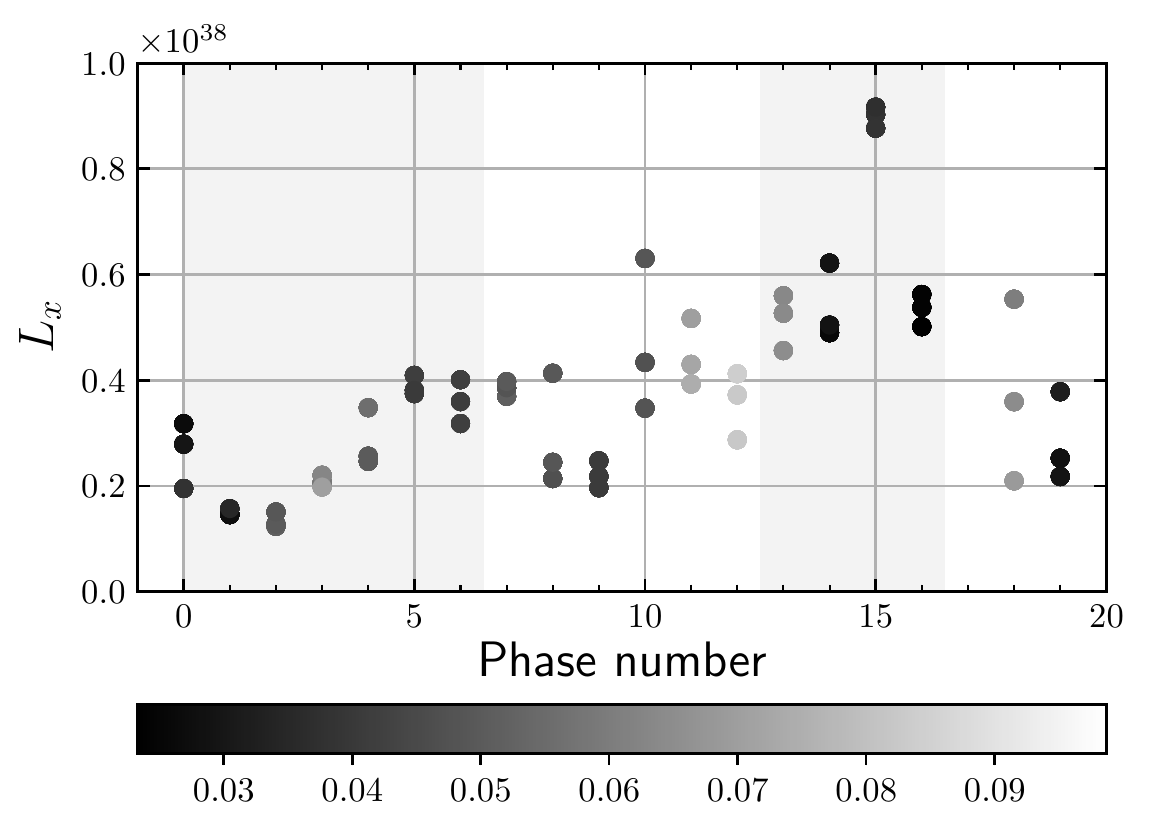}
    \caption{Total X-ray luminosity of the neutron star for the models described in Fig. \ref{f:theta_Z}.}
    \label{f:Lx}
\end{figure}

In Fig.~\ref{f:Lx} we plot the total X-ray luminosity of the neutron star as a function of the 35-day phase. The meaning of the filled circles is the same as in the two previous Figures. The spread in the X-ray luminosity found from the best-fitting of the optical light curves can be due to the construction of the observed light curves from different 35-day cycles. The total X-ray luminosity clearly demonstrates the growth from the main-on to short-on state. Physically, this may be connected with the storage of matter in the disk during the main-on when the optical star is mostly illuminated by the X-ray source and the gas stream through the Lagrange point $L_1$ from the optical star is the most powerful. The time delay between the main-on and short-on states roughly corresponds to the viscous time of the accretion disk \citep{Klochkov2006}.  

\subsubsection{Outer and inner disk viewing angles}

\begin{figure}
  \centering
  \includegraphics[width=0.4\textwidth]{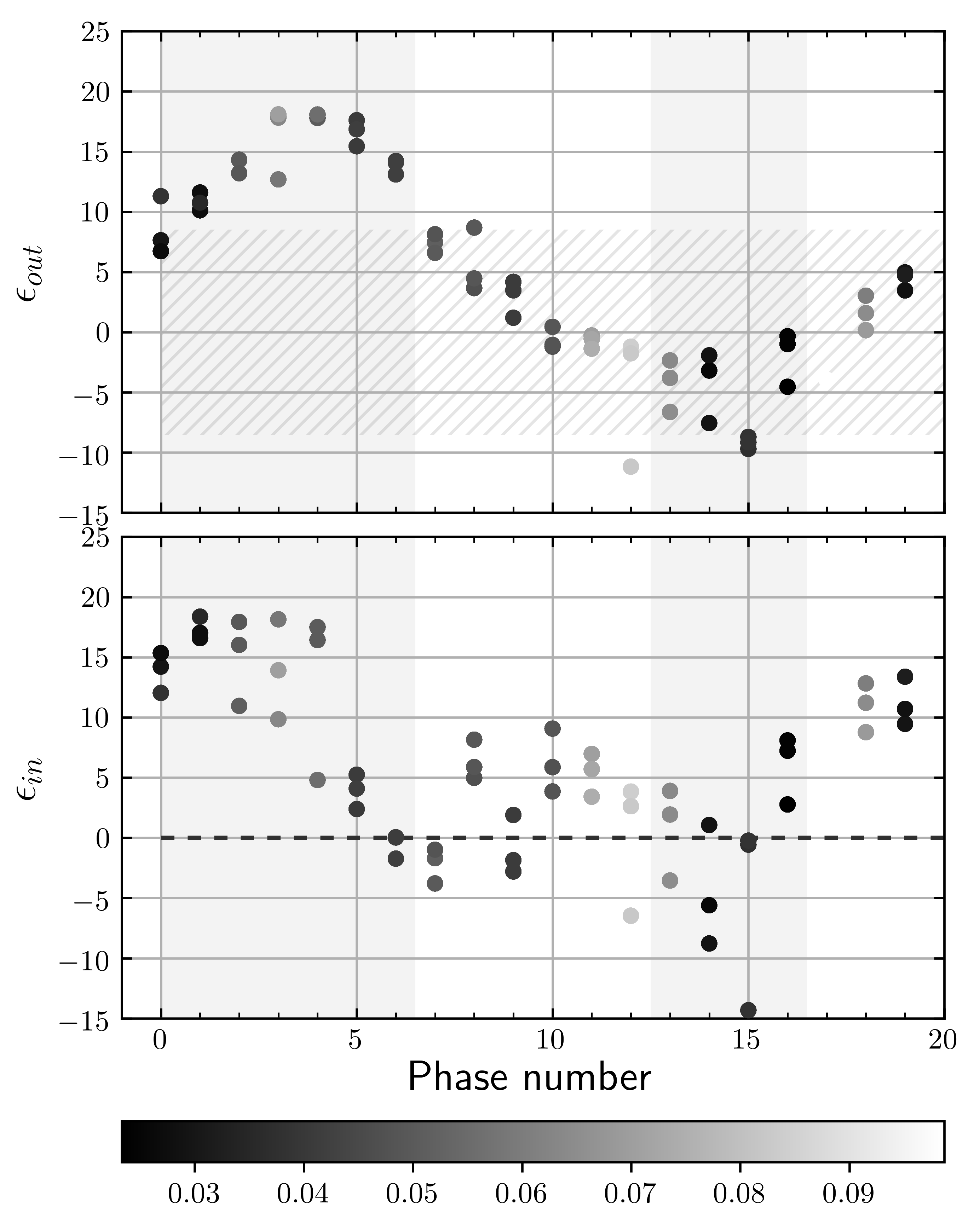}
  \caption{The angle between the outer and inner disk edge with the viewing angle ($\epsilon_{out}$ and $\epsilon_{in}$, the upper and bottom panels, respectively) for the models from Fig. \ref{f:theta_Z}. Hatched area in the upper graph indicate the area where the X-ray radiation is blocked by the outer edge of the disk.
    }
  \label{f:epsilon}
\end{figure}

In Fig. \ref{f:epsilon} we show the angles between the outer ($\epsilon_{out}$) and inner ($\epsilon_{in}$) disk planes and the viewing angle (the upper and bottom panels, respectively). In the upper panel, the light hatched strip marks the range of $\epsilon_{out}$ inside which the X-ray source is screened by the outer disk plane (the low states of Her X-1). It is seen that $\epsilon_{out}$ behaves with the 35-day phase in a way enabling the main-on state. As for the short-on, several points appears inside the screened area, which means that no X-ray radiation should be visible by observer. The visible disagreement with the obtained best-fit $\epsilon_{out}<H/R\sim-8\degree$ can be related to the different 35-day cycles used to construct the optical light curves and to the likely variability of the disk thickness with the 35-day phase. In the bottom panel, the angle $\epsilon_{in}$ vanishes by the end of the main-on, which is indeed expected because the X-ray spectroscopic observations \citep{2005A&A...443..753K} suggest that the hot inner parts of the accretion disk screens the X-ray source at the main-on termination. 

\subsubsection{Outer disk retrograde precession}

\begin{figure}
  \centering
  \includegraphics[width=0.4\textwidth]{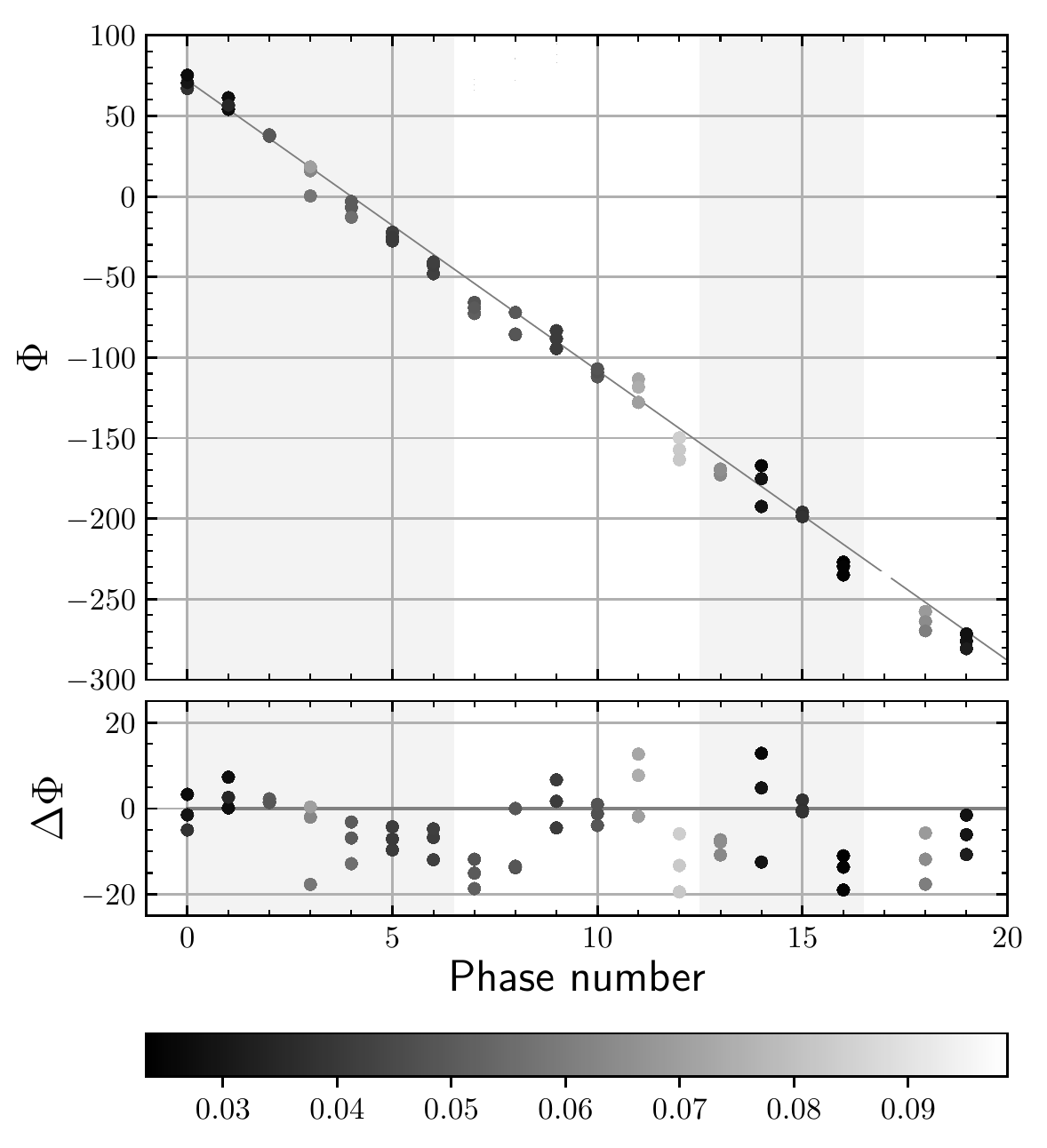}
  \caption{The phase angle of the outer disk edge $\Phi$ as a function of the 35-day phase interval $n$ (the upper panel) and its dispersion $\Delta \Phi$ relative to the linear law (the bottom panel) for the same models as in Fig. \ref{f:theta_Z}.}
  \label{f:phi}
\end{figure}

Fig. \ref{f:phi} shows the phase angle of the outer disk with the 35-day phase (the upper panel). Clearly, this plot confirms our assumption about an almost even rate of the outer disk retrograde precession. Deviations in the disk phase angle $\Delta \Phi$ from the linear law $\Phi=\Omega t$ are within the narrow range $\pm 20\degree$ (the bottom panel), which may be either due to physical variability or a large collection of different 35-day cycles used to construct the optical light curves.

\section{Discussion}

The results presented in the present paper support the physical model of the 35-day cycle of HZ~Her/Her~X-1 based on the free precession of the neutron star and retrograde precession of the surrounding twisted warped accretion disk 
\citep{1990ESASP.311..235P, 2000astro.ph.10035K, 2009A&A...494.1025S, 2013MNRAS.435.1147P}. In this model, the synchronization between the neutron star free precession period and the precession period of the accretion disc is mediated by variable gas stream from the Lagrange point $L_1$. The properties of the stream are modulated with the neutron star free precession period because of variable X-ray illumination of the optical star atmosphere. 

In the present study we have made several assumptions to be discussed.
First, we have fixed the binary parameters of HZ Her/Her X-1, the neutron star parameters (viewing angle, location of emitting regions on the surface around magnetic poles, X-ray emission diagram, etc.), accretion disk radius and outer disk thickness, which were adopted from previous works (see Table \ref{tab:parameters1}). Of them, the neutron star parameters could mostly affect the synthesized light curves because they determine the X-ray illumination of the optical star atmosphere.

Second, we have ignored the detailed temperature distribution across the accretion disk, which may be quite complicated. Therefore, we have not modelled orbital phases 0.0--0.13 and 0.87--1.0 during which eclipsing effects by the optical star are important.
The binary inclination $i$ is known to be close to $90\degree$. In our modelling, we computed it from the following conditions: 1) the outer disk precession is uniform; 2) the outer disk thickness is constant; 3) the X-ray source opens by the outer disk edge both in main-on and short-on states; 4) the maximum disk opening occurs at the precession phase angle $\Phi_0 = 2\pi/5$. Clearly, either of these conditions are approximate, but the change of $i$ by $1\degree$ doesn't change qualitative conclusions of our modelling. 

We also discuss a possible syncronisation mechanism between the neutron star free precesion and accretion disc precession.

The observed periodic change of the X-ray pulse shape favours the neutron star free precession. In this model, the neutron star free precession period is very close to that of the accretion disk precession, suggesting a synchronization  mechanism between the neutron star and accretion disk motion.

The inner part of the accretion disk is warped by the magnetic torque of the neutron star magnetosphere. The torque changes during the neutron star free precession period which affects the character and degree of the curvature of the inner disk parts.

Dynamic action of accretion streams interacting with the disk tends to increase the disk precession period. Without the dynamic action of accretion streams, the disk precession period would be determined by the tidal torques only and would be shorter than observed. We suggest that the neutron star -- disk synchronization is possible if the periods of the neutron star free precession and accretion disk are close to each other.

However, the neutron star crust is subject to cracking because of variable torques. The neutron star crust crack can abruptly change the neutron star free precession period, see equation \ref{e:Omegap}. If a crack is large enough, synchronization can break, decreasing the anisotropy of the disk shadow on the optical star atmosphere and hence the outer accretion disk tilt to the orbital plane. In this case, an anomalous low state of Her X-1 where the X-ray emission is fully blocked by the accretion disk can occur. However, during the anomalous low state the inner part of the accretion disk subjected to the magnetic torque from the neutron star magnetosphere keeps warped, blocking the X-ray emission from the neutron star in some directions, and the disk shadow on the optical star surface still remains asymmetrical relative to the orbital plane. Therefore, the emerging gas pressure gradients near the inner Lagrangian point will deflect the accretion stream from the orbital plane. Ultimately, such deflected accretion streams may tilt the outer parts of the disk again, enabling the opening of the X-ray source for the observer during the disk precessional motion.

According to our modelling, the magnetic dipole axis far from the surface of the neutron star (at the distance $\sim 100\,R_{ns}$) should be misaligned with the direction to the north magnetic pole on the neutron star surface by about $10\degree$--$15\degree$ (see Fig.~\ref{f:NS_free_precession}). It may be due to a complex structure of the magnetic field near the neutron star surface, as suggested by the modelling of X-ray pulse profiles \citep{2013MNRAS.435.1147P}.

In our model, the neutron star X-ray luminosity changes by a factor of $\sim 3$ with the maximum being at the short-on phase. During the short-on phase, the north magnetic pole of the freely precessing neutrons star is closest to the neutron star spin axis. It means that at these phases most of the X-ray radiation is directed away from the optical star and the observer. This is why the increase in $L_x$ at these phases does not affect both the optical light curve shape and the observable X-ray flux.

Our model suggests that during the main-on state, where the maximal X-ray irradiation of the optical star occurs, the matter additionally accumulates in the disk. This excess matter accretes onto the neutron star on a viscous time scale of the disk, which should be of the order of many orbital periods.

%\textcolor{blue}{Note that the optical light curves of HZ Her/Her X-1 could be reproduced in our model with twisted precessing accretion disk and an isotropic pulse-averaged central X-ray source with adjustable X-ray luminosity at each 35-day cycle phase. However, in this case it would be difficult to explain the observed strong changes in X-ray pulse profiles, see} \citep{2013MNRAS.435.1147P}.

%Our modelling shows that the introduction of a complex X-ray beam shape from the neutron star (as suggested by the X-ray pulse profile modeling \citep{2013MNRAS.435.1147P}) decreases $\chi^2$.
%\replaced{A model with an isotropic central X-ray source fits the data with worse reduced $\chi^2$ (see Fig.\ref{f:chi2})}{However, even a simpler model with an isotropic X-ray beam from the neutron star is able to adequately reproduce the observed optical light curves.}

\section{Conclusions}

In this paper, we calculated orbital light curves of HZ~Her/Her~X-1 for 20 phases of 35-day superorbital cycle. The main features of the model include a tilted, warped and precessing accretion disk and a freely precessing neutron star. The precessing accretion disk produces a complex varying shadow on the atmosphere of the optical star shaping the optical light curve. 
The freely precessing neutron star serves as the clock mechanism providing the long-term stability of the 35-day cycle. 
We find that the model with a warped precessing tilted disk can adequately reproduce the observed light curves photometrical behaviour under physically motivated choice of the model parameters.

The precession angle of the disk linearly increase with the 35-day phase with a maximum opening of the X-ray source to the observer during the main-on state of Her X-1 at the phase $\simeq 0.2$. Geometrical parameters of the disk vary with the 35-day cycle. Only inner regions of the disk are warped due to the magnetic torque from the neutron star magnetosphere. The warp magnitude and sign are in agreement with the predicted behaviour of the magnetic torque as a function of the angle between the magnetic dipole axis and the neutron star spin axis which changes periodically with the neutron star 
free precession phase.

%\textbf{The model with a complex form of the X-ray radiation beam (as suggested by the X-ray pulse profile modelling assuming the neutron star free precession) yields smaller $\chi^2$ than the model with isotropic X-ray radiation from the neutron star. The first one also is in agreement with statement of free precession of the neutron star.}

The model parameters depending on the 35-day phase change smoothly. The spread of some parameters is due to the use of observations taken over long period of time covering a large number of 35-day cycles.

The obtained high X-ray luminosity of the source during the short-on phase is likely to be due to accumulation of  matter in the outer disk during the main-on and latter accretion of the accumulated matter in the viscous disk time. During the short-on state, the north magnetic pole of the neutron star has the largest viewing angle and directed away from the optical star. Therefore, no significant changes in the optical light curves amplitude and X-ray flux are observed.

We discuss the possible synchronization mechanism between the neutron star free precession and the disk precession.
It can be related to the dynamical interaction of gas streams from the inner Lagrangian point with the outer parts of the disk. The streams can flow out of the orbital plane because of the uneven X-ray illumination of the optical star atmosphere produced by the changing shadow of the tilted twisted accretion disk. In this model, the anomalous low states of Her X-1 could be a consequence of crust glitches of the freely precessing neutron star. Such a glitch changes abruptly the neutron star precession period and hence breaks the neutron star -- accretion disk synchronization. This results in a more even illumination of the optical star atmosphere and decrease of the disk tilt. In the absence of the neutron star free precession the recovery of the disk tilt and reappearance of the X-ray source would be impossible.

\section*{Acknowledgements}
The research is supported by the RFBR grant No.\,18-502-12025 (carrying out of the observations and data processing), the DFG grant No. 259364563 (processing of the X-ray data) and program of Leading  Science School MSU (Physics of Stars, Relativistic Compact Objects and Galaxies). N. Shakura acknowledges a partial support by the Russian Government Program of Competitive Growth of Kazan Federal University. I. Bikmaev and E. Irtuganov thank TUBITAK, IKI, KFU, and AST for partial support in using RTT150 (the Russian-Turkish 1.5-m telescope in Antalya). The work of I. Bikmaev and E. Irtuganov was partially funded by the subsidy 671-2020-0052 allocated to Kazan Federal University for the state assignment in the sphere of scientic activities. The observations of I. Volkov were fulfilled with 1-m reflector of Simeiz observatory of INASAN. The work of I. Volkov was partially supported by a scholarship of Slovak Academic Information Agency SAIA. The work of S. Yu. Shugarov was supported by the Slovak Academy of Sciences grant VEGA No. 2/0008/17, by the Slovak Research and Development Agency under the contract No. APVV-15-0458.

\section*{Data Availability}
The data and code underlying this article are available in the GitHub Repository at \url{https://github.com/eliseys/data} and \url{https://github.com/eliseys/discostar} respectively.

%%%%%%%%%%%%%%%%%%%%%%%%%%%%%%%%%%%%%%%%%%%%%%%%%%



\bibliographystyle{mnras}
\bibliography{mybib}

% \begin{thebibliography}{99}
% \bibitem[\protect\citeauthoryear{Author}{2012}]{Author2012}
% Author A.~N., 2013, Journal of Improbable Astronomy, 1, 1
% \bibitem[\protect\citeauthoryear{Others}{2013}]{Others2013}
% Others S., 2012, Journal of Interesting Stuff, 17, 198
% \end{thebibliography}

%%%%%%%%%%%%%%%%%%%%%%%%%%%%%%%%%%%%%%%%%%%%%%%%%%

%%%%%%%%%%%%%%%%% APPENDICES %%%%%%%%%%%%%%%%%%%%%

% Don't change these lines
\bsp	% typesetting comment
\label{lastpage}
\end{document}